\documentclass[conference]{IEEEtran}

\usepackage{amssymb}
\usepackage[ruled,vlined,linesnumbered]{algorithm2e}
\usepackage{amsthm}
\usepackage{amsfonts}
\usepackage{amsmath}
\usepackage{booktabs}
\usepackage{bm}
\usepackage{color}
\usepackage{cite}
\usepackage{comment}
\usepackage{epsfig}
\usepackage{euscript}
\usepackage{fancybox}
\usepackage{graphicx}
\usepackage{hyperref}
\usepackage{lipsum}
\usepackage{pgf}
\usepackage{psfrag}
\usepackage{pifont}
\usepackage{mathrsfs}
\usepackage{multirow}
\usepackage{setspace}
\usepackage{stfloats}
\usepackage{tabularx}
\usepackage{textcomp}
\usepackage[caption=false, font=footnotesize]{subfig}

\usepackage[anythingbreaks]{breakurl}

\newcolumntype{C}[1]{>{\centering\arraybackslash}p{#1}}
    \def\Complex{{\rm\rule[.23ex]{.03em}{1.1ex}\kern-.3em{C}}}

    \newcommand{\be}{\begin{equation}} \newcommand{\ee}{\end{equation}}
    \newcommand{\bea}{\begin{eqnarray}} \newcommand{\eea}{\end{eqnarray}}
    \newcommand{\benum}{\begin{enumerate}} \newcommand{\eenum}{\end{enumerate}}

    \newcommand*{\argmin}{\operatornamewithlimits{argmin}\limits}

\begin{document}

\title{Edge Intelligence Empowered UAVs for Automated Wind Farm Monitoring in Smart Grids}

\author{\IEEEauthorblockN{
Hwei-Ming Chung\IEEEauthorrefmark{1}, Sabita Maharjan\IEEEauthorrefmark{1}\IEEEauthorrefmark{2}, Yan Zhang\IEEEauthorrefmark{1}\IEEEauthorrefmark{2}, Frank Eliassen\IEEEauthorrefmark{1}, and Tingting Yuan\IEEEauthorrefmark{3}}

\IEEEauthorblockA{\IEEEauthorrefmark{1} Department of Informatics, University of Oslo, 0373 Oslo, Norway}
\IEEEauthorblockA{\IEEEauthorrefmark{2} Simula Metropolitan Center for Digital Engineering, 0167 Oslo, Norway}
\IEEEauthorblockA{\IEEEauthorrefmark{3} Institute of Computer Science, Georg-August-University of G{\"o}ttingen, 37073 G{\"o}ttingen, Germany}

\IEEEauthorblockA{Emails: \IEEEauthorrefmark{1}hweiminc@ifi.uio.no, sabita@ifi.uio.no, yanzhang@ieee.org, frank@ifi.uio.no
\IEEEauthorrefmark{3}tingt.yuan@hotmail.com 
}

\vspace{-0.35cm} 

}

%\markboth{IEEE Communications Letters,~Vol.~XX, No.~XX, XXX~2013}%
%{Shell \MakeLowercase{\textit{et al.}}: Bare Demo of IEEEtran.cls for Journals}

\maketitle
\begin{abstract}

%Wind power is regarded as a promising choice of alternative energy for the future energy system.
With the exploitation of wind power, more turbines will be deployed at remote areas possibly with harsh working conditions (e.g., offshore wind farm).
The adverse working environment may lead to massive operating and maintenance costs of turbines.
Deploying unmanned aerial vehicles (UAVs) for turbine inspection is considered as a viable alternative to manual inspections.
An important objective of automated UAV inspection is to minimize the flight time of the UAVs to inspect all the turbines. 
A first contribution of this paper is thus formulating an optimization problem to compute the optimal routes for turbine inspection satisfying the above goal.
On the other hand, the limited computational capability on UAVs can be used to increase the power generation of wind turbine.
Power generation from the turbines can be optimized by controlling the yaw angle of the turbines. 
%To solve the problems, we require the forecasted wind information.
Forecasting wind conditions such as wind speed and wind direction is crucial for solving both optimization problems.
Therefore, UAVs can utilize their limited computational capability to perform wind forecasting.
In this way, UAVs form edge intelligence in offshore wind farm.
%Specifically, sensors embedded in UAVs for turbine inspection can be applied to remote sensing.
%With high-resolution measurements from remote sensing, UAVs can utilize limited computational capability to perform wind forecasting such that UAVs can form edge intelligence in offshore wind farm.
% With the measurements from the remote sensing, UAVs can further act as computing units (i.e., mobile edge computing servers) to perform turbine control.
%Therefore, we forecast the wind information while considering the computational limit of UAVs.
With the forecasted wind conditions, we design two algorithms to solve the formulated problems, and then evaluate the proposed methods with real-world data.
The results reveal that the proposed methods offer an improvement of $44\%$ of the power generation from the turbine compared to hour-ahead forecasting and $25\%$ reduction of the flight time of the UAVs compared to the chosen baseline method.

\end{abstract}
\textbf{\textit{Index terms--} unmanned aerial vehicle (UAV), edge intelligence, wind forecast, turbine control, routing problem.}

\section{Introduction}
\IEEEPARstart{W}{ind} power is a clean alternative to address climate change and to reduce our dependence on fossil fuel power.
The total capacity of the installed wind power in Europe is expected to reach $350$ GW by $2030$, which can support $24\%$ of the European electricity consumption \cite{europe_wind}.
%Moreover, wind power is expected to support $24\%$ of the electricity demand by $2030$.
Also, grid operators can operate power grids with flexibility by utilizing wind power so as to form a smart grid.
This goal requires more wind turbines to be installed.
However, turbines may suffer from failure of components, such as blades, gearbox, and yaw system \cite{2007-turbine-failure}, which can affect the efficiency of the turbine and can also lead to compete operation failure. 
For example, a blade failure can result in a downtime of more than seven days \cite{2014-DTU-report}.

Currently, the inspection of wind turbines is done manually by qualified service personnel.
This procedure can take from several days to several weeks, requiring intensive and costly efforts.
Moreover, the workers are exposed to variable wind conditions and harsh environments in offshore wind farms.
To reduce the utilization of manpower, unmanned aerial vehicles (UAVs) can play a crucial role in automating the inspection of wind turbines. 
For instance, cracks on the surface of turbine blades can be detected with the help of images taken by UAVs \cite{2017-UAV-blade-wang,2019-UAV-blade-wang}.
%The authors in \cite{2019-UAV-inspec-Hoang}  for the UAV fleet to inspect the wind turbines.
Additionally, UAVs can be embedded with various advanced sensors such as Lidar sensors \cite{2016-lidar-inspect-UAV}, thermal-wave radar \cite{2018-blade-thermal-yang}, and millimeter wave imaging \cite{2017-blade-mmwave-wang} to detect the failure of wind turbines.
With advanced sensors, automated inspection with UAVs has also been studied in \cite{2015-UAV-PV-panel,2018-UAV-power-inspect}.
In \cite{2015-UAV-PV-panel}, UAVs with thermal-wave radar were utilized to monitor the condition of solar panels.
The inspection of power lines using UAVs was proposed in \cite{2018-UAV-power-inspect} so that workers do not need to climb the transmission line tower.
In the context of the smart grid, automatic meter reading is another application of UAVs in power systems \cite{2017-power-reading-UAV}.
UAVs combined with fault indicators was proposed in \cite{2019-UAV-power-inspect} to deliver signal from the fault indicators to the operator in case of damage on the distribution line.

UAVs have also obtained some computational capabilities and therefore UAVs can also be applied to facilitate computing service at remote sites.
That is, computation can be moved closer to the edge of the network (i.e., UAVs) such that data can be processed physically close to where the data is generated.
In \cite{2019-UAV-MEC-route}, optimal UAV-routes were computed for UAV assisted inspection of wind turbines under different wind conditions.
That is, UAVs can observe current wind condition and then computational resources of the UAVs can be utilized to calculate the optimal routing paths for inspection.
%Also, UAVs can help deliver data from sensors to the operator through cellular network or satellite.  
%This framework has also been applied in wireless communication \cite{2019-UAV-MEC-offload,2020-UAV-lstm-task-pre}.
%In \cite{2018-Du-UAV-MEC}, the authors applied UAVs to collect the data from the sensors and then the resource of UAVs is allocated based on different quality of service (QoS) requirement of sensors. 
%{\rl Then, UAVs can act as temporary mobile base stations such that the stress of base station can be reduced as stated in \cite{2019-UAV-MEC-wireless-mag}.}
%In \cite{2019-UAV-MEC-offload}, UAVs solved an optimization problem for minimizing the total energy consumption of UAVs when UAVs were deployed to offload computing tasks from mobile devices.
More recently, combining machine learning algorithms with the computational resources of the UAVs has become a promising approach towards managing more complicated tasks, thus giving rise to the term "edge intelligence".
%More recently, the use of machine learning algorithms can continuously improve the performance of the computing tasks in the edge devices.
%In addition, utilizing the computational resources of UAVs has become a promising approach towards managing more complicated tasks, thus giving rise to the term "edge intelligence".
UAVs were utilized to schedule the transmission of the signal for users to prevent from jamming by applying deep reinforcement learning \cite{2019-UAV-DRL-intellegence}.
\cite{2020-UAV-lstm-task-pre} is a recent work in this context where the authors proposed a framework where UAVs utilized long short term memory (LSTM) to predict the upcoming tasks and then to find the optimal position to serve mobile users.

In previous works, UAVs were mainly regarded as relays in \cite{2019-UAV-DRL-intellegence,2020-UAV-lstm-task-pre} and as inspection tools in \cite{2015-UAV-PV-panel,2017-power-reading-UAV,2018-UAV-power-inspect,2019-UAV-power-inspect}.
However, sensors embedded in UAVs can be utilized for remote sensing \cite{2018-UAV-sensing}.
That is, Lidar sensors \cite{2016-lidar-inspect-UAV} can be applied to collect the meteorological measurements in the wind farm.
UAVs acting as computing units can provide information (e.g., wind speed and wind direction) to wind turbines thus playing a significant role in controlling the operation of wind turbines.
Specifically, it is expected that wind farms become larger and larger in the near future while the capacity of turbines is also increasing.
Small variation of wind conditions in large wind farms considerably affect the power generation of wind turbines with high capacity.
In this case, UAVs can obtain high-resolution meteorological measurements from remote sensing and then forecast weather conditions to turbines, so that the wind turbines can operate at their optimal conditions to generate maximum power. 
Moreover, the forecasting information supplies valuable information for the decision-making of the routes for turbine inspection.
For computation necessary for forecasting however, it is required to consider the limited computational capabilities and storage space on UAVs.

In this paper, we utilize UAVs to form the edge intelligence units in an offshore wind farm.
That is, we utilize UAVs as remote sensing units to collect the meteorological measurements in offshore wind farms.
With the measurements, UAVs can inspect the turbines and can also control the operation of wind turbines by adjusting the yaw angle.
In this regard, we formulate a control optimization problem to maximize power generation from turbines and an optimization problem to find the optimal routing path for automated turbine inspection by UAVs.
To solve the problems, UAVs forecast the wind speed and wind direction based on the measurements collected by sensors embedded in the UAV.
Moreover, we take into account the hardware limitation of the UAVs for forecasting wind speed and wind direction based on machine learning method.
Then, we design two algorithms to solve the formulated problems by making use of the forecasting results as inputs.
To this end, our main contributions in this paper are threefold:
\begin{itemize}
\item We present a new framework for utilizing UAVs as edge intelligence units to facilitate automated wind farm monitoring.
That is, we formulate an optimization problem for maximizing the power generation of turbines and a UAV-route optimization problem for turbine inspection.

\item We design heuristic algorithms to derive the optimal condition for operating wind turbines and the optimal path for the inspection with forecasted wind as input.
The results reveal that $25\%$ reduction on the flight time compared to the baseline method and $44\%$ increase in power generation compared to hour-ahead forecasting.

\item UAVs have some computational capabilities and storage space.
Therefore, we take the limited computational capabilities and storage space on UAVs into account when forecasting the wind speed and wind direction by LSTM.

\end{itemize}

\section{System Model}\label{sec:system_model_problem_formulation}

\subsection{Wind and Wind Turbine Model}

We denote the wind velocity by $\mathbf{w} = [ w^{x}, w^{y} ]$, where $w^{x}$ and $w^{y}$ represent the projection of the wind velocity on the x-axis and the y-axis, respectively.
The wind speed, denoted by $w_{s}$, is calculated from $w_{s} =||\mathbf{w}||_{2}$.
Let $\theta_{w}^{pol}$ be the wind direction in the polar coordinate system defined as $ \theta_{w}^{pol} = \arctan \frac{w^{y}}{w^{x}}$.
The representation of wind direction in the meteorological measurements is different from $ \theta_{w}^{pol}$.
In this case, we define the wind direction in the meteorological measurements as $\theta_{w}^{met}$.
Phases $0$, $\frac{\pi}{2}$, $\pi$, and $\frac{3 \pi}{2}$ are used to represent the north, the east, the south, and the west wind in the meteorological measurements, respectively.
With this notion, the phase is represented in a clockwise direction.
In the polar coordinate system however, the phase is represented in a counterclockwise direction.
Therefore, $\theta_{w}^{pol} $ and $\theta_{w}^{met}$ are related as $\theta_{w}^{pol} = \frac{3 \pi}{2} - \theta_{w}^{met}$.

\begin{figure}
\begin{center}
\resizebox{1.5in}{!}{%
\includegraphics*{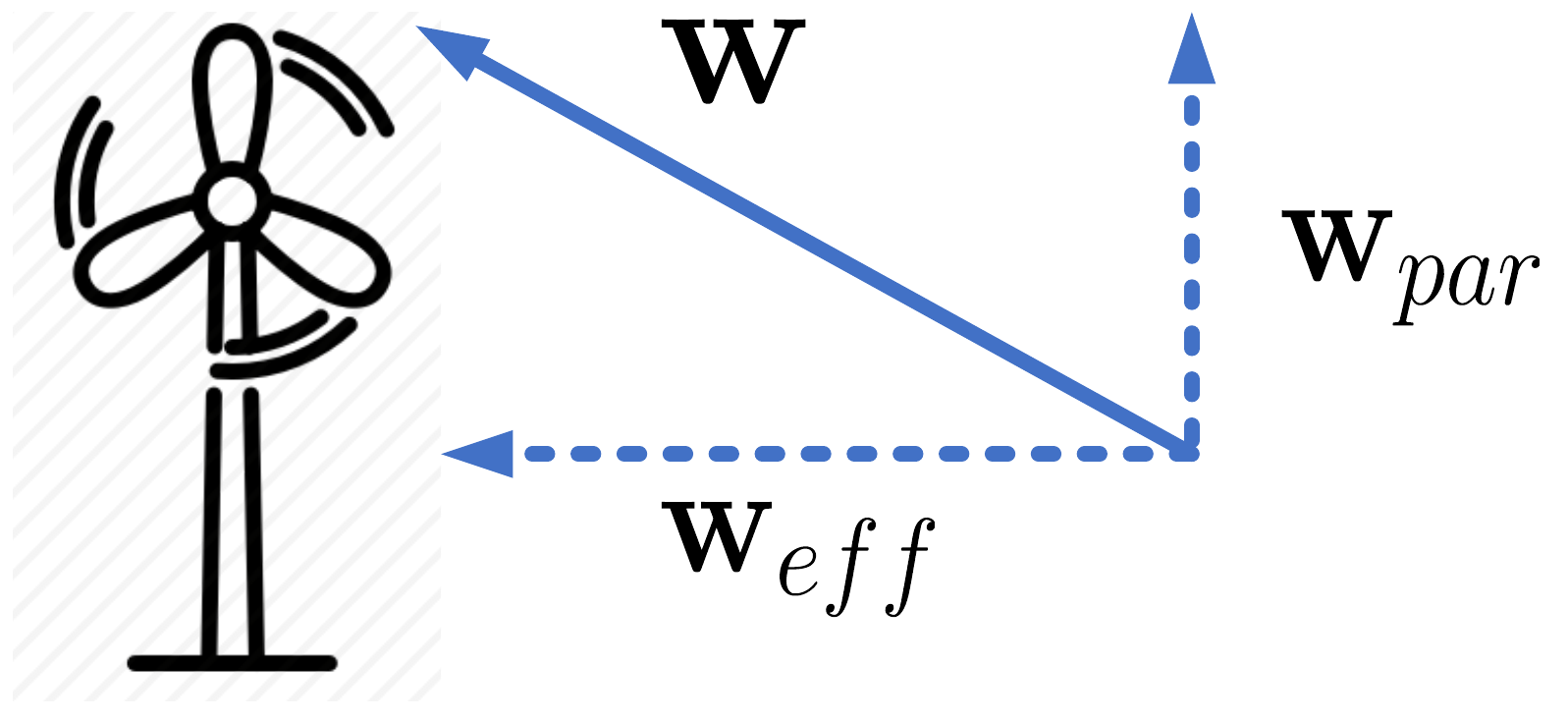} }%
\vspace{-0.4cm} 
\caption{The relation between a wind turbine and the wind velocity.} 
\vspace{-0.4cm} 
\label{fig:turbine_wind_model}
\end{center}
\end{figure}

The coordinates of the $k$-th turbine in the wind farm are $\mathbf{q}_{k} = [x_{k}, y_{k}]$.
Then, according to \cite{2014-cc-power-curve} and \cite{2014-ks-power-curve}, the power generation of the $k$-th turbine can be represented as
\vspace{-0.15cm}
\begin{equation} \label{eq:power_gene_turbine}
p_{k} = \frac{1}{2} \rho A w_{s}^{3} C_{p}(\lambda, \beta),
\vspace{-0.15cm}
\end{equation}
where $A$ is the blade sweep area and $\rho$ is the air density.
Function $C_{p}$ represents the power coefficient of the turbine based on $\lambda$ and $\beta$.
Quantities $\lambda$ and $\beta$ are the pitch angle of wind turbine and the tip speed ratio, respectively.
However, for a wind turbine, the power generation is only affected by the wind velocity that is perpendicular to the turbine.
Therefore, $w_{s}$ in (\ref{eq:power_gene_turbine}) should be replaced.
Wind velocity can be separated into two components, $\mathbf{w}_{eff}$ and $\mathbf{w}_{par}$, which respectively represent the wind direction perpendicular and parallel to the turbines.
In this case, the power generation of turbines is only affected by $\mathbf{w}_{eff}$.
We then replace $w_{s}$ in (\ref{eq:power_gene_turbine}) with $||\mathbf{w}_{eff}||_{2}$.
The relation between a turbine and wind velocity is shown in Fig. \ref{fig:turbine_wind_model}.

\subsection{UAV Model}\label{subsec:UAV_model}

We consider a total of $N$ UAVs in an offshore wind farm to inspect the condition of the turbines.
Each UAV $i=1, ...N$, is placed at $\mathbf{q}_{i} = [ x_{i}, y_{i} ]$.
The set of turbines for UAV $i$ to inspect is denoted by ${\cal N}_{i}$, and $|{\cal N}_{i}|$ represents the cardinality of set ${\cal N}_{i}$.

\begin{figure}
\begin{center}
\resizebox{2.4in}{!}{%
\includegraphics*{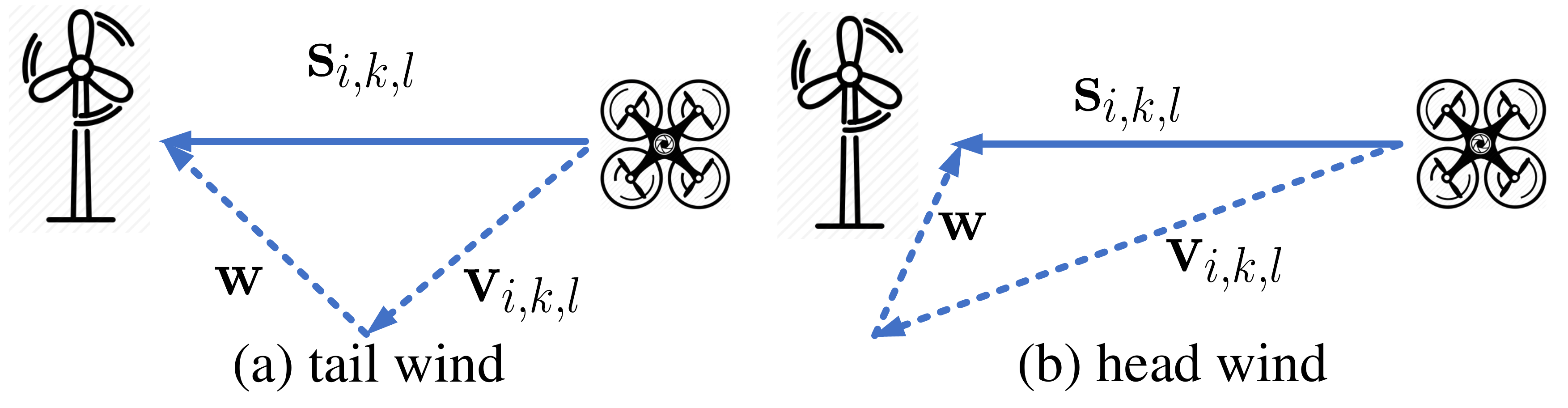} }%
\vspace{-0.4cm} 
\caption{The relation between a UAV, a wind turbine, and the wind velocity.} 
\vspace{-0.4cm} 
\label{fig:UAV_wind_model}
\end{center}
\end{figure}

When inspecting the turbines, wind speed and wind direction must be considered for deriving the optimal UAV-route for turbine inspection.
Thus, UAV $i$ may face two wind conditions, namely tail wind and head wind, as shown in Fig. \ref{fig:UAV_wind_model}.
We define $\mathbf{s}_{i, k, l}$ and $\mathbf{v}_{i, k, l}$ as the resultant velocity and the UAV velocity of UAV $i$ flying from turbine $k$ to turbine $l$, respectively.
Vector $\mathbf{v}_{i, k, l}$ is the initial velocity of the UAV, and the resultant velocity is defined by the velocity influenced by the wind velocity.
The angle between $\mathbf{s}_{i, k, l}$ and $\mathbf{v}_{i, k, l}$ is denoted by $\theta_{i, k, l}^{s, v}$ and $\theta_{i, k, l}^{s, w}$ is defined as the angle between $\mathbf{s}_{i, k, l}$ and $\mathbf{w}$.
The relation between $\mathbf{v}_{i, k, l}$, $\mathbf{s}_{i, k, l}$, and $\mathbf{w}$ is expressed as
\vspace{-0.25cm}
\begin{equation} \label{eq:velocity-relation}
\mathbf{v}_{i, k, l} + \mathbf{w} = \mathbf{s}_{i, k, l}.
\vspace{-0.2cm}
\end{equation}
We then define the airspeed and the groundspeed as $||\mathbf{v}_{i, k, l}||_{2}$ and $||\mathbf{s}_{i, k, l}||_{2}$, respectively.
The flying speed limit is denoted by $u_{i}^{max}$.
% For each UAV, it has the speed limit of $u_{i}^{max}$ and the maximum wind speed resistance is denoted by $u_{i}^{wind}$
% For the wind situation in Fig. \ref{fig:UAV_wind_model}, groundspeed is limited to $u_{i}^{max}$ in Fig. \ref{fig:UAV_wind_model}(a) and the groundspeed is limited to $u_{i}^{max}$ in Fig. \ref{fig:UAV_wind_model}(b).
% Then, the maximum wind speed resistance is denoted by $u_{i}^{wind}$.

The flight time for UAV $i$ traveling from turbine $k$ to turbine $l$ can be calculated as
\vspace{-0.1cm}
\begin{equation} \label{eq:travel_time}
t_{i, k, l} = \frac{ ||\mathbf{q}_{l} - \mathbf{q}_{k}||_{2} }{||\mathbf{s}_{i, k, l}||_{2}}.
\vspace{-0.1cm}
\end{equation}
The maximum flight time of UAV $i$ is denoted by $t_{i}^{max}$, which represents the upper limit on the total flight time during inspection.
%The distance between UAV $i$ and UAV $j$ is denoted by $d_{i, j}$, and the distance between UAV $i$ and turbine $k$ is represented as $d_{i, k}$. 
The distance between UAV $i$ and turbine $k$ is $d_{i, k}$.

%The flying range of UAV $i$ under wind conditions $\mathbf{w}$ can be expressed as 
The flying range of UAV $i$ can be expressed as 
\vspace{-0.2cm}
\begin{equation} \label{eq:flying_range}
B_{i} (\rho_{i}) =  \{ x, y \in \mathbb{R}: ||\mathbf{r}||_{2} \leq \rho_{i} \},
\vspace{-0.2cm}
\end{equation}
where $\mathbf{r} = [ x-x_{i}, y - y_{i} ]$.
%\begin{equation}
%\mathbf{r} = [ x-x_{i}, y - y_{i} ].
%\end{equation}
The quantity $\rho_{i}$ is the actual flying distance of UAV $i$, which is defined by the operator.
As in (\ref{eq:flying_range}), we regard the flying range of the UAV as a circle, with the location of UAV $i$, $[ x_{i}, y_{i} ]$ as the center of the circle.

\section{Problem Formulation}\label{sec:problem_formulation}

In this section, we formulate a control optimization problem to obtain the maximum power generation from the wind turbines.
Then, we formulate a routing optimization problem to find the best UAV path for the inspection.
Wind is a very important factor in determining the flying range and flying speed of UAVs.
We therefore also incorporate the influence of the wind speed and the wind direction into the routing optimization problem.

\subsection{Power Generation Maximization with Yaw Control}

Let $\theta_{k}^{yaw}$ be the yaw angle that can be adjusted to generate the maximum power generation of the $k$-th turbine
Let $\theta_{k}^{w, yaw}$ denote the angle between $\mathbf{w}$ and $\theta_{k}^{yaw}$.
With these notations, an optimization problem for maximizing the power generation of the turbine can be formulated as 
\begingroup
\allowdisplaybreaks
\begin{subequations}\label{eq:max_power_gene}
\begin{align}
\max_{\theta_{k}^{yaw}}  &  ~~ p_{k}  \\
\mbox{subject to}    & ~~ 0 \leq \theta_{k}^{yaw} \leq 2\pi , \label{eq:angle_limit} \\
& ~~ 0 \leq p_{k} \leq p_{k}^{max},  \label{eq:gene_limit} \\
& ~~ ||\mathbf{w}_{eff}||_{2} = w_{s} \cos(\theta_{k}^{w, yaw}).  \label{eq:w_eff_gene} 
\end{align}
\end{subequations}
\endgroup
%The objective function of (\ref{eq:max_power_gene}) is to maximize the power generation of the turbine.
The yaw angle of the turbine is limited between $0$ and $2\pi$ as indicated in (\ref{eq:angle_limit}).
Constraint (\ref{eq:gene_limit}) implies that the power generation of turbines cannot exceed the rated power, $p_{k}^{max}$.
The power generation is affected by the wind direction perpendicular to the turbine, $||\mathbf{w}_{eff}||_{2}$, which is given by (\ref{eq:w_eff_gene}).

\subsection{Routing Optimization of UAVs}

\begin{comment}
In the wind farm, we route the UAVs to inspect the wind turbines.
Then, we can use a graph, denoted by $\mathcal{G}_{i}=\{ {\cal N}_{i}, {\cal E}_{i}$\}, to represent the relation of UAV $i$ and the turbines assigned to it.
Set ${\cal N}_{i}$ denotes the nodes in the graph and the edges between the turbines in  ${\cal N}_{i}$ is mentioned in ${\cal E}_{i}$.
We can create an adjacency matrix, represented as $\mathbf{D}_{i}$, based on the graph structure.
The element of the $k$-th column and the $l$-th row in $\mathbf{D}_{i}$ is $t_{i, k, l}$, which is the flight time from turbine $k$ to $l$.
The value of $t_{i, k, l}$ and the value of $t_{i, l, k}$ are not the same because of the wind, and therefore the adjacency matrix is not symmetric.
Thus, $\mathcal{G}_{i}$ is an asymmetric graph. 
\end{comment}

%In the wind farm, we route the UAVs to inspect the wind turbines.
Let $M_{i}$ denote the number of required routes for UAV $i$ to inspect the turbines.
We introduce matrix $\mathbf{U}_{i}^{m} = [U_{i, k, l}^{m}]_{|{\cal N}_{i}|\times |{\cal N}_{i}|}$ to denote the $m$-th route for UAV $i$.
If UAV $i$ chooses to fly from turbine $k$ to turbine $l$, $U_{i, k, l}^{m}$ is $1$; otherwise, $U_{i, k, l}^{m}$ is $0$.
The optimal routing problem can then be formulated as
\begingroup
\allowdisplaybreaks
%\vspace{-0.1cm}
\begin{subequations}\label{eq:routing_problem}
\begin{align}
\min_{\substack{ M_{i}, \mathbf{U}_{i}^{m}, \mathbf{v}_{i, k, l},\\ \mathbf{s}_{i, k, l}, \theta_{i, k, l}^{s, v}}}  &  ~ t^{ins} = \sum_{i} \sum_{m=1}^{M_{i}}\sum_{k \in {\cal N}_{i} } \sum_{l \in {\cal N}_{i} \setminus \{ k\} } t_{i, k, l} U_{i, k, l}^{m}   \\
\mbox{subject to}    & ~~ \sum_{k \in {\cal N}_{i}} U_{i, s, k}^{m} = \sum_{k \in {\cal N}_{i}} U_{i, k, s}^{m} = 1, \quad \forall m, i \label{eq:start_point} \\
                     & \sum_{l \in {\cal N}_{i} \setminus \{ l\}  } U_{i, l, k}^{m} = \!\!\!\sum_{l \in {\cal N}_{i} \setminus \{ k\}  } U_{i, k, l}^{m} = 1, \forall m, i\label{eq:link_point} \\
                     & ~~ U_{i, k, l}^{m} \in \{ 0, 1\}, \quad\quad\qquad \forall k, l \in  {\cal N}_{i}, \forall i \label{eq:binary_limit} \\
                     &  \sum_{k \in Q} \sum_{l \in Q} U_{i, k, l}^{m} \leq |Q|\!-\! 1,   \!\forall Q  \subsetneq  {\cal N}_{i}, m, \forall i \label{eq:avoid_subtour} \\
                     & ~~ 1 \leq M \leq |{\cal N}_{i}|-1,    \qquad \qquad \qquad \quad  \forall i \label{eq:route_limit}\\
                     & ~~ \sum_{k, l \in {\cal N}_{i}} t_{i, k, l} U_{i, k, l}^{m} \leq t_{i}^{max}, \quad\qquad\forall m, i \label{eq:time_limit}\\
                     & ~~ ||\mathbf{v}_{i, k, l}||_{2}\leq u_{i}^{max},  \quad\qquad \forall k, l \in  {\cal N}_{i} , \forall i \label{eq:airspeed_limit}\\
                     & ~~ ||\mathbf{s}_{i, k, l}||_{2}\leq u_{i}^{max},  ~\quad\qquad \forall k, l \in  {\cal N}_{i} , \forall i \label{eq:ground_limit}\\
                     & ~~ \mathbf{v}_{i, k, l} + \mathbf{w} = \mathbf{s}_{i, k, l},    \quad \qquad \forall k, l \in  {\cal N}_{i} ,\forall i \label{eq:wind_speed_relation}
\end{align}
\end{subequations}
\endgroup
In (\ref{eq:routing_problem}), the objective is to minimize the flight time and the number of routes for all UAVs in the wind farm to inspect all the turbines.
The location of the UAV is regarded as a starting point, $s = \mathbf{q}_{i}$, as indicated in (\ref{eq:start_point}).
Then, only one route can be obtained between turbines as stated in (\ref{eq:link_point}).
The path of routing during inspection, $U_{i, k, l}^{m}$, is restricted to a binary parameter in (\ref{eq:binary_limit}).
Constraint (\ref{eq:avoid_subtour}) ensures that a closed path does not exist in the subset $Q$ of ${\cal N}_{i}$.
The number of routes for the inspection should be less than the number of turbines in ${\cal N}_{i}$ as stated in (\ref{eq:route_limit}).
For every route, the total flight time cannot exceed $t_{i}^{max}$ according to (\ref{eq:time_limit}).
The airspeed and groundspeed should be bounded by $u_{i}^{max}$ as mentioned in (\ref{eq:airspeed_limit}) and (\ref{eq:ground_limit}), respectively.
The relationship between the wind velocity, the UAV velocity, and the resultant velocity mentioned in (\ref{eq:velocity-relation}) is given by (\ref{eq:wind_speed_relation}).

\section{Algorithm Design}\label{sec:forecast_model}

To design our algorithms, we require the future wind condition (i.e., wind speed and wind direction) as input.
UAVs acting as remote sensing units in wind farms can obtain meteorological measurements with high resolution.
In this case, these measurements can be applied to perform wind forecasting.
The UAVs have limited computational capabilities and storage space, and therefore we need to consider the hardware limitation when performing wind forecasting.

\subsection{Wind Forecasting with Low-precision LSTM}\label{dubsec:q_LSTM}

%We introduce LSTM to UAVs to forecast the wind such that UAVs can provide wind information to enhance the operation of wind turbines.
%However, UAVs do not have enough storage space and computational resource to handle the complex computation of wind forecasting.
%Thus, we intend to reduce the usage of the computational resource and the storage space for the wind forecasting.
We introduce LSTM to UAVs to forecast the wind speed and the wind direction, and the structure of LSTM can be described as 
\begingroup
\allowdisplaybreaks
\begin{subequations}\label{eq:LSTM_eq}
\begin{align}
f_{i} & = \sigma_{g} (\mathbf{W}_{f} \mathbf{x}_{t}  + \mathbf{U}_{f} \mathbf{h}_{t-1} + \mathbf{b}_{f}), \\
i_{i} & = \sigma_{g} (\mathbf{W}_{i} \mathbf{x}_{t}  + \mathbf{U}_{i} \mathbf{h}_{t-1} + \mathbf{b}_{i}), \\
o_{t} & = \sigma_{g} (\mathbf{W}_{o} \mathbf{x}_{t}  + \mathbf{U}_{o} \mathbf{h}_{t-1} + \mathbf{b}_{o}), \\
c_{t} & = f_{t} \circ \mathbf{c}_{t-1} + i_{t} \circ \sigma_{c} (\mathbf{W}_{c} \mathbf{x}_{t}  + \mathbf{U}_{c} \mathbf{h}_{t-1} + \mathbf{b}_{c}), \\
h_{t} & = o_{t} \circ \sigma_{h} (\mathbf{c}_{t}), 
\end{align}
\end{subequations}
\endgroup
where $\circ$ denotes the Hadamard product.
The input data is denoted by $\mathbf{x}_{t}$.
In (\ref{eq:LSTM_eq}), $f_{i}$, $i_{i}$, $o_{t}$, $c_{t}$, and $h_{t}$ are the forget gate, the input gate, the output gate, the cell state, and the hidden state, respectively.
Then, $\mathbf{b}_{f}$, $\mathbf{b}_{i}$, $\mathbf{b}_{o}$, and $\mathbf{b}_{c}$ are the bias vectors.
To address the limitation of the hardware, we introduce the quantization to LSTM.
We use $\omega_{w}$ to denote the number of bits of fixed-point integer to represent all elements in $\mathbf{W}_{f}$, $\mathbf{W}_{i}$, $\mathbf{W}_{o}$, $\mathbf{W}_{c}$, $\mathbf{U}_{f}$, $\mathbf{U}_{i}$, $\mathbf{U}_{o}$, and $\mathbf{U}_{c}$.
The quantization model is represented as 
\vspace{-0.15cm}
\begin{equation}
Q_{\omega_{w}} (\mathbf{W}) = clip \left( \frac{ \mathbf{W} }{ \gamma \mbox{median}(|\mathbf{W}|) }  , -0.5, 0.5 \right) + 0.5, 
\vspace{-0.15cm}
\end{equation}
where a natural choice of $\gamma$ would be $2.5$ \cite{2016-qrnn-he}, and the clip function is defined as 
\vspace{-0.15cm}
\begin{equation}
clip \left( \mathbf{W} \right) = \frac{1}{ 2^{\omega_{w}} - 1 }  \left \lfloor (2^{\omega_{w}} - 1) \mathbf{W} + 0.5 \right \rfloor,
\vspace{-0.15cm}
\end{equation}
where $\lfloor * \rfloor$ is the round function.
In the final layer, we introduce a fully-connected layer to obtain the forecasted value, $\hat{y}_{t}$, as 
\vspace{-0.15cm}
\begin{equation}\label{eq:LSTM_fc}
\hat{\mathbf{y}}_{t} = \mathbf{W}_{y} h_{t} + \mathbf{b}_{y}.
\vspace{-0.15cm}
\end{equation}
In (\ref{eq:LSTM_fc}), $\mathbf{W}_{y}$ is quantized with $\omega_{f}$ bits and $\mathbf{b}_{y}$ is the corresponding bias vector.

With the quantized LSTM framework, we can further reduce the computational complexity for wind forecasting.
According to \cite{2016-ICLR-quan}, the weight matrix becomes sparse with quantization.
Therefore, we further utilize the sparse matrix multiplication to the proposed framework.

\subsection{Algorithms for Maximizing Power Generation}

With the forecasted wind conditions, we can now solve problem (\ref{eq:max_power_gene}).
Before calculating power generation, UAVs calculate the power coefficient for the turbines (the optimal working condition for the turbines).
Then, we obtain the power generation of the turbine by (\ref{eq:power_gene_turbine}).
We change the yaw angle to $\theta_{w}^{pol}$ when $p_{k}$ is lower than the rated power.
Adjusting the yaw angle is stopped when the power generation reaches the rated power.
The details are provided in Algorithm \ref{ago:yaw-control}.
Then, the computational complexity of Algorithm \ref{ago:yaw-control} is ${\cal O}( |{\cal N}_{i}| )$, where it is only related to the number of turbines for UAV $i$ to inspect.

\begin{algorithm}
%\small
\caption{Search the Optimal Yaw Angle}
\label{ago:yaw-control}
\DontPrintSemicolon
\KwIn{$\hat{\mathbf{y}}_{t}$, $\rho$, $A$, $\theta_{w}^{pol}$}
\KwOut{$\theta_{k}^{yaw}$}
%Initialize the parameters \\
Obtain $ C_{p}(\lambda, \beta)$ by using the method in \cite{2014-cc-power-curve}\;
Compute the power generation with (\ref{eq:power_gene_turbine}) \;
$
\theta_{k}^{yaw} =
\left\{
\begin{array}{ll}
                \theta_{k}^{yaw} = \theta_{w}^{pol}, \quad   & \mbox{if} ~ p_{k} \leq p_{k}^{max},    \\
                \theta_{k}^{yaw} = \theta_{k}^{yaw}, \quad   & \mbox{if} ~ p_{k} > p_{k}^{max}.
\end{array}
        \right.
$

\end{algorithm}

\subsection{Algorithms for Finding Optimal Routing Path}

\begin{algorithm}
%\small
\caption{Obtain the optimal routing path}
\label{ago:optimal-route}
 \DontPrintSemicolon
\KwIn{ ${\cal N}_{i}$, $[x_{i}, y_{i}]$, $B_{i}$, $\hat{\mathbf{y}}_{t}$}
\KwOut{$U_{i, k, l}^{m}$}
%Initialize the parameters \\
Set $\hat{\cal{N}}_{i} = {\cal N}_{i}$; $s= [x_{i}, y_{i}]$; \;
Calculate the flying range based on $\hat{\mathbf{y}}_{t}$ \;
Perform the intersection of $B_{i}^{\hat{\mathbf{y}}_{t}}$ and $B_{i}$ to get $Z_{i}$ \;
Put the turbines out of new flying range to $o_{i}$ \;
Sort $o_{i}$ of all UAVs based on $|o_{i}|$ in a decreasing order with the index as $e_1,e_2,\dots ,e_{N}$\;
\For{$i = 1 $ \KwTo $ N$}{
\While{$o_{e_{i}} > 0$}{
	Take turbine $k$ from $o_{e_{i}}$ \;
	Assign turbine $k$ to UAV $j$ according to (\ref{eq:reassign_turbine}) \;
	${\cal N}_{i}  = {\cal N}_{i} \setminus \{ k \} $; $o_{e_{i}} = o_{e_{i}} \!\setminus\! \{ k \}$; $\hat{\cal{N}}_{j}  = \hat{\cal{N}}_{j} \cup \{ k \} $ \;
}

}
Calculate $t_{i, k, l}$ with $k, l \in \hat{\cal{N}}_{j}$ by (\ref{eq:travel_time}) and (\ref{eq:s_veloci_cal}) \;
Use Branch-and-Cut algorithm to find the optimal path without considering (\ref{eq:time_limit}) \;
Separate route based on $t_{i}^{max}$ to get $U_{i, k, l}^{m}$ for UAV $i$ \;
\end{algorithm}

With the forecasted wind speed and wind direction, we can now obtain the solution of (\ref{eq:routing_problem}).
The solving procedure is presented in Algorithm \ref{ago:optimal-route}.
At the beginning, the UAV calculates the flying range based on forecasted wind conditions, denoted by $B_{i}^{\hat{\mathbf{y}}_{t}}$.
The center of $B_{i}^{\hat{\mathbf{y}}_{t}}$, $[ \hat{x_{r}}, \hat{y_{r}} ]$, is defined as
\vspace{-0.15cm}
\begin{equation} \label{eq:center_point_cal}
\hat{x_{r}} = x_{i} + w^{x} t_{i}^{max}, ~~ \hat{y_{r}} = y_{i} + w^{y} t_{i}^{max}.  
\vspace{-0.15cm}
\end{equation}
The new flying range of the UAV is the intersection of $B_{i}^{\hat{\mathbf{y}}_{t}}$ and $B_{i}$ denoted by
\vspace{-0.15cm}
\begin{equation}\label{eq:flying_range_intersec}
Z_{i} = B_{i}^{\hat{\mathbf{y}}_{t}} \bigcap B_{i}.
\vspace{-0.15cm}
\end{equation}
With the new flying range, some turbines in ${\cal N}_{i}$ may beyond the range $Z_{i}$.
The set of turbines outside $Z_{i}$ is denoted by $o_{i}$.
Then, we sort $|o_{i}|$ of all UAVs in decreasing order.
We assign turbine $k$ in $o_{i}$ to UAV $j$ based on 
\vspace{-0.15cm}
\begin{equation}\label{eq:reassign_turbine}
j = \left\{  \argmin_{j} d_{j, k} | [ x_{k}, y_{k} ] \in Z_{j} \right\}.
\vspace{-0.15cm}
\end{equation}
The reassigned set of turbine to UAV $i$ is denoted by $\hat{\cal{N}}_{i}$.
Then, we use (\ref{eq:travel_time}) to calculate $t_{i, k, l}$.
Vector $\mathbf{s}_{i, k, l} $ is obtained from
\begin{equation} \label{eq:s_veloci_cal}
\!\!\!\!\mathbf{s}_{i, k, l} \!\!=\!\! \left\{
\begin{array}{lll}
                \left[u_{i}^{max} \cos(\theta_{s}), u_{i}^{max} \sin(\theta_{s}) \right], &  \!\!\!\! 0 \leq \theta_{i, k, l}^{s, w} \leq \frac{\pi}{2}, \\
                \left[ u_{i}^{s} \cos(\theta_{s}), u_{i}^{s} \sin(\theta_{s}) \right], & \!\!\!\!\frac{\pi}{2} < \theta_{i, k, l}^{s, w} \leq \pi,
\end{array}
        \right.\!\!
\end{equation}
where $\theta_{s}$ is given by $\arctan((y_{l}-y_{k})/(x_{l}-x_{k}))$.
In (\ref{eq:s_veloci_cal}), $u_{i}^{s}$ can be calculated using $ u_{i}^{s} = u_{i}^{max} \cos (\theta_{i, k, l}^{s, v}) - w_{s} \cos (\pi - \theta_{i, k, l}^{s, w})$, where $\theta_{i, k, l}^{s, v} = \arcsin ( w_{s} \sin (\pi - \theta_{i, k, l}^{s, w}) /  u_{i}^{max} )$.
Then, we can use the relation in (\ref{eq:velocity-relation}) to calculate $\mathbf{v}_{i, k, l}$.
Finally, line $12$ in Algorithm \ref{ago:optimal-route} (Algorithm $2$ in \cite{2019-UAV-route-optimize-fuel}) was applied to find the optimal routing path without considering flight time constraint, (\ref{eq:time_limit}).
Then, we separate the optimal path by considering $t_{i}^{max}$ at line $13$ in Algorithm \ref{ago:optimal-route} (Algorithm $3$ in \cite{2019-UAV-route-optimize-fuel}).
The computational complexity of Algorithm \ref{ago:optimal-route} is ${\cal O} \left( N \log(N) + N^{2}|o_{i}| + |\hat{\cal{N}}_{i}|^{2} 2^{|\hat{\cal{N}}_{i}|} + |\hat{\cal{N}}_{i}| \right) $.

\section{Numerical Results}\label{sec:simulation}

In this section, we evaluate the performance of the proposed method based on a real-world dataset.
We use the offshore wind data at Roland island recorded by National Renewable Energy Laboratory (NREL) \cite{wind_data_NREL}.
This dataset obtained from \cite{wind_data_NREL} is referred to as 5-min-wind, which contains the wind speed and direction with $5$-min resolution.
Dataset 5-min-wind can be regarded as the measurements collected by the remote sensing with UAVs.
Traditionally, the resolution of the wind data for forecasting is 1 hr.
Therefore, we construct the wind data with 1-hour resolution from 5-min-wind, referred to as 1-hr-wind, for comparison.
Then, forecasting the wind with 1-hr-wind is referred to as hour-ahead forecasting.

The make of the UAVs used in the simulation is AscTec Falcon 8.
It is embedded with Lidar sensors to inspect turbines.
%\footnote{http://www.asctec.de/en/uav-uas-drones-rpas-roav/asctec-falcon-8/\#pane-0-1}.
%This UAV can carry diverse sensors, namely Lidar, ultrasonic sensor, and camera, to inspect turbines.
It has a flight time between $12$ and $22$ minutes; so we set $t_{i}^{max}$ to $18$ minutes.
The airspeed is limited to $16$ m/s.
The maximum resistance of the UAV to the wind speed is $15$ m/s.
The placement of UAVs in the wind farm and the turbines assigned to UAVs are solved by K-means clustering and non-linear integer programming \cite{2019-UAV-place}.

The make of the turbine used in the simulation is SG 8.0-167 DD from Siemens.
%\footnote{https://www.siemensgamesa.com/products-and-services/offshore/wind-turbine-sg-8-0-167-dd}.
The swipe area, $A$, is $21900$ square meter.
The rated power is $8$ MW and $\lambda$ is set to $0^{\circ}$.
The tip speed ratio is generated according to the relation mentioned in \cite{2014-cc-power-curve}.
The air density, $\rho$, is set to $1.065 \mbox{kg}/\mbox{m}^{3}$.

For wind forecasting, we utilize the wind data from previous $2$ hrs to forecast the wind data for next $40$ mins.
In the LSTM, it has a hidden layer with $100$ nodes and is implemented in TensorFlow $1.13$ with the Python $3.7.7$.
We compare the mean absolute error (MAE) and root mean square error (RMSE) as defined by (\ref{eq:MAE_RMSE}) for performance comparison.
\vspace{-0.15cm}
\begin{equation} \label{eq:MAE_RMSE}
\!\!\!\!\! MAE =  \sum_{i=1}^{n}  \frac{ | y_{i} - \hat{y}_{i} | }{n}, RMSE = \sqrt{ \sum_{i=1}^{n} \frac{  ( y_{i} - \hat{y}_{i})^{2}  }{n} }.
\vspace{-0.15cm}
\end{equation}
\begin{comment}
\begingroup
\allowdisplaybreaks
\begin{subequations}\label{eq:routing_problem}
\begin{align}
MAE =  & ~~  \sum_{i=1}^{n}  \frac{ | y_{i} - \hat{y}_{i} | }{n}, \label{eq:MAE_def} \\
RMSE = & ~~ \sqrt{ \sum_{i=1}^{n} \frac{  ( y_{i} - \hat{y}_{i})^{2}  }{n} }. \label{eq:RMSE_def} 
\end{align}
\end{subequations}
\endgroup
\end{comment}
In the simulation, MAE is used as the loss function of the LSTM network.

\subsection{Wind Forecasting Result}

For solving both optimization problems, we utilize LSTM for forecasting wind speed and direction. 
We first compare the performance of wind forecasting by utilizing remote sensing with traditional hour-ahead forecasting.
Then, we introduce quantization to the LSTM.
We compare the accuracy with and without quantization and find out how many bits are enough for wind forecasting.

The forecasting results with dataset 5-min-wind and 1-hr-wind are provided in Table \ref{tb:accuracy_compare_du}.
According to the results, the MAE for forecasting the wind speed and the wind direction with hour-ahead forecasting is $0.0719$ and $0.1051$, respectively.
However, the wind forecast with remote sensing yields $43\%$ reduction on MAE compared to the hour-ahead forecasting.
The improvement in terms of RMSE shows comparable numbers.
This implies that remote sensing can considerably improve the accuracy of wind forecasting.

\begin{table} \footnotesize
\begin{center}
\caption{Forecasting accuracy under different dataset}
\label{tb:accuracy_compare_du}
\vspace{-0.3cm}
\begin{tabular}{|c|c|c|c|c|c|c|c|c|}
\hline
		&  \multicolumn{2}{c|}{Wind Speed}			& \multicolumn{2}{c|}{Wind Direction} \\
\hline 
Method   & MAE        & RMSE       & MAE       & RMSE   \\
\hline
Remote Sensing 	& $0.0403$      & $ 0.0573 $  		& $0.0596$      & $ 0.1151 $   \\
\hline
Hour Ahead 	& $0.0719$       & $ 0.0972 $       & $0.1051$      & $ 0.1687 $  \\
\hline

\end{tabular}
\vspace{-0.5cm}
\end{center}
\end{table}

\begin{comment}
\begin{table} \small
\begin{center}
\caption{Reward Comparison}
\label{tb:accracy_diff_float}
\begin{tabular}{|c|c|c|c|c|c|c|c|c|}
\hline
		&  \multicolumn{2}{c|}{Wind Speed}			& \multicolumn{2}{c|}{Wind Direction} \\
\hline
   & MAE     & RMSE       & MAE    & RMSE   \\
\hline
float32 	& $0.0403$      & $ 0.0573 $  	  & $0.0596$      & $ 0.1151 $   \\
\hline
float16 	& $0.0396$      & $ 0.0554 $      & $0.0609  $    & $ 0.1192$  \\
\hline

\end{tabular}
\end{center}
\end{table}
\end{comment}

%In the previous result, we can ensure that the remote sensing with UAVs can bring higher accuracy.
The accuracy of forecasting wind speed and wind direction under different quantization is shown in Table \ref{tb:accracy_diff_bit}.
For the first row, $\omega_{w}$ and $\omega_{f}$ with both float32 ($32$-bit floating point) represent wind forecasting without quantization.
Then, we compare the scenario where weights are quantized with $16$-bit, $8$-bit, $4$-bit, and $2$-bit fixed-point integer.
%and computing wind forecasting with $16$-bit floating point.
%From $16$-bit floating point to $8$-bit fixed point, the MAE and RMSE vary slightly. 
According to the results, we can obtain similar accuracy as the forecasting without quantization if we further quantize the weight to $4$ bits.
In this case, we only require one-eighth storage space compared to the conventional LSTM using $32$-bit floating point.
%In the later simulation, we use the results of wind forecasting with $4$ bits as the input.

\begin{table} \footnotesize
\begin{center}
\caption{Forecasting Accuracy under Different Bits for Quantization}
\label{tb:accracy_diff_bit}
\vspace{-0.3cm}
\begin{tabular}{|c|c|c|c|c|c|c|c|c|}
\hline
\multicolumn{2}{|c|}{}		&  \multicolumn{2}{c|}{Wind Speed}			& \multicolumn{2}{c|}{Wind Direction} \\
\hline
$\omega_{w}$ & $\omega_{f}$ & MAE     & RMSE       & MAE    & RMSE   \\
\hline
float32 	& float32 & $0.0403$      & $ 0.0573 $  	  & $0.0596$      & $ 0.1151 $   \\
\hline
16  & 16	& $0.0396$      & $ 0.0554 $       & $0.0609  $    & $ 0.1192$  \\
\hline
%8  & float16	& $0.0365$      & $ 0.0525 $      & $0.0603 $    & $0.1177$  \\
%\hline
8  & 8	& $0.0392$      & $ 0.0579 $      & $ 0.0608 $    & $ 0.1191 $  \\
\hline
4  & 4	& $0.0401$      & $ 0.0571 $      & $0.0598 $    & $ 0.1183 $  \\
\hline
2  & 2	& $0.0431$      & $ 0.0614 $      & $ 0.0807 $    & $ 0.1308 $  \\
\hline

\end{tabular}
\vspace{-0.5cm}
\end{center}
\end{table}

\subsection{Power Generation with Yaw Control}

We apply the results of wind forecasting as input to the yaw control to compare the power generation.
In this section, we compare power generation of a wind turbine for an hour and for a day.
We input the real wind data, the forecasting results using remote sensing, and forecasting results of using hour-ahead forecasting to Algorithm \ref{ago:yaw-control} and then compare the power generation in Table \ref{tb:power_gene_yaw}.
In Table \ref{tb:power_gene_yaw}, we list the forecasted angle and then calculate the corresponding power generation for an hour.
We can observe that we lose $2$ kW of power generation in an hour if we forecast wind with remote sensing.
However, a huge difference between forecasted and real wind direction occurs when utilizing hour-ahead forecasting.
The incorrect wind direction can lead to non-optimal power generation from the turbines which means considerable loss in terms of huge loss on the power generation.
We observe that with our proposed framework, we can obtain $44.45\%$ increase in power generation in a day if we forecast the wind with remote sensing compared to hour-ahead forecasting.
%In this context, we can conclude that the proposed framework can benefit the power generation of wind turbine.

\begin{table} \footnotesize
\begin{center}
\caption{The wind power generation under different wind data}
\label{tb:power_gene_yaw}
\vspace{-0.3cm}
\begin{tabular}{|c|c|c|c|c|c|c|c|}
\hline
\multirow{2}{*}{ \shortstack{Method} }   & \multirow{2}{*}{ $\theta_{w}^{pol}$(deg.) }   &  \multirow{2}{*}{ \shortstack{Generation in\\an hour (kWh)} }\\
& &  \\ 
 \hline \hline
True                   & $10$   & $129$           \\
\hline \hline
Remote Sensing         & $16$   & $127$           \\
\hline \hline
Hour Ahead             & $49$   & $44$           \\
\hline 

\end{tabular}
\begin{tabular}{|c|c|c|c|c|c|c|c|}
\hline
 \multirow{2}{*}{ \shortstack{Generation in\\a day (kWh)} }  \\ \\
\hline \hline
$19861.2077$  \\
\hline \hline
$17952.1253$  \\
\hline \hline
$12428.0077$     \\
\hline 

\end{tabular}
\vspace{-0.5cm}
\end{center}
\end{table}

\subsection{Routing with Results of Wind Forecasting}

\begin{figure}
\begin{center}
\resizebox{2.8in}{!}{%
\includegraphics*{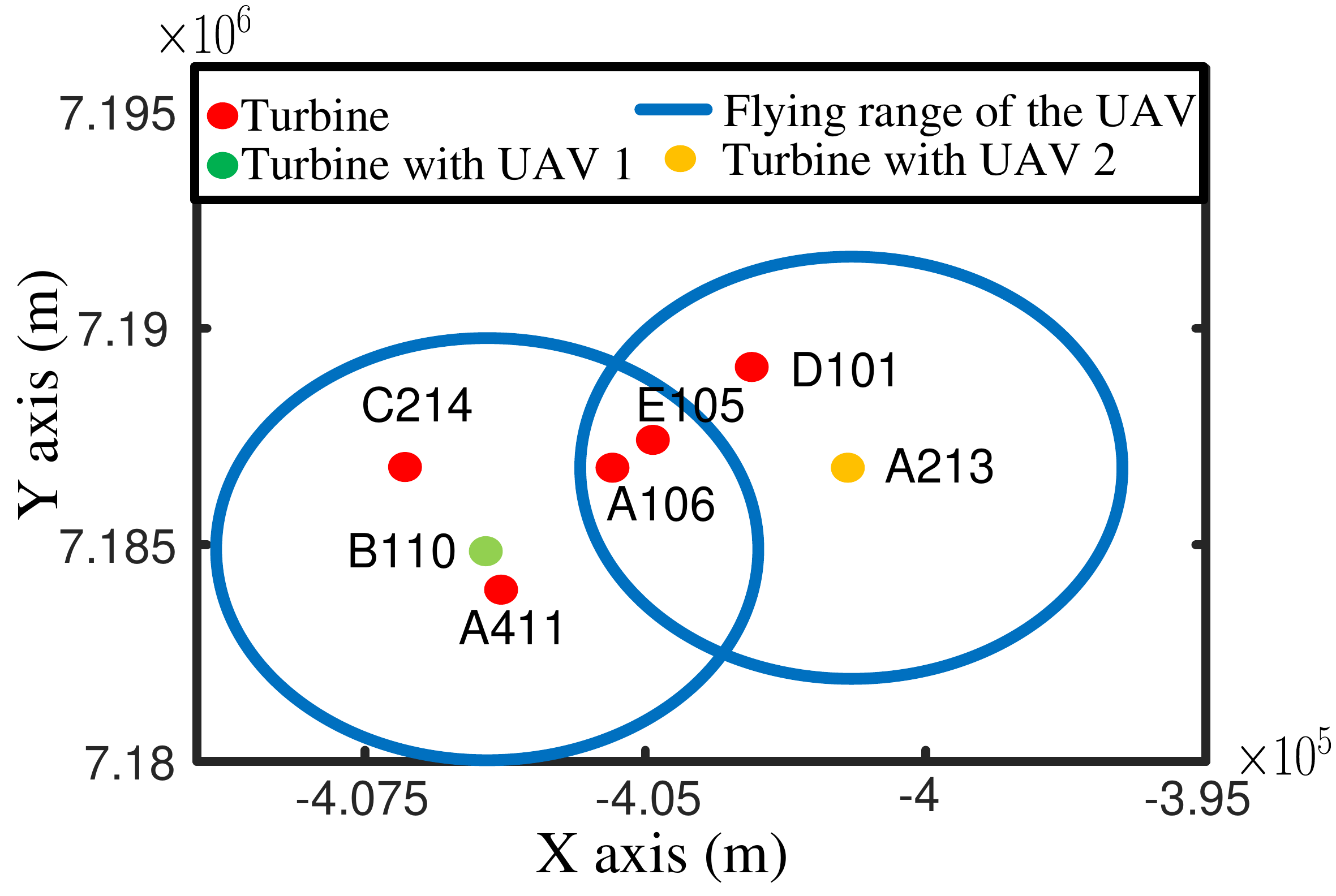} }%
\vspace{-0.4cm} 
\caption{The coordinates of UAVs and turbines.} 
\vspace{-0.5cm} 
\label{fig:UAV_range}
\end{center}
\end{figure}

Finally, we input the forecasted wind data to Algorithm \ref{ago:optimal-route} to solve the routing problem.
We compare Algorithm \ref{ago:optimal-route} with the algorithms for finding the optimal routing paths in \cite{2019-UAV-route-optimize-fuel}.
A diagram showing all UAVs and all turbines may lack clarity, and therefore we provide the results with two UAVs here.
The location of the UAVs and the turbines assigned to them are presented in Fig. \ref{fig:UAV_range}.
In Fig. \ref{fig:UAV_range}, UAV $1$ and $2$ are responsible for inspecting $5$ and $2$ turbines, respectively.
Then, we consider a wind condition with $w_{s} = 10$ m/s and $\theta_{w}^{met}$ to $\frac{\pi}{2}$ (east wind).
The routing results are summarized in Table \ref{tb:route_result_dynamic}.

In Table \ref{tb:route_result_dynamic}, we compare the routing results with and without using Algorithm \ref{ago:optimal-route}.
The total inspection time without using the proposed algorithm is $40.5051$ mins.
This is because UAV $1$ faces head wind when flying to inspect turbine E105 and A106.
Thus, UAV $1$ may leave these $2$ turbines for the next round.
Going back to the starting point and then starting another round for inspection takes extra time.
By using Algorithm \ref{ago:optimal-route}, E105 is outside the flying range of UAV $1$.
In this case, E105 is assigned to UAV $2$ for inspection.
If the Algorithm \ref{ago:optimal-route} is utilized, we can temporarily assign E105 to UAV $2$ under this wind condition.
By doing so, both UAVs only require one round to finish the inspection.
Moreover, the total inspection time is reduced to $30.0741$ mins, which is equivalent to $25\%$ reduction in inspection time.

\begin{table} \footnotesize
\begin{center}
\caption{The routing results with and without dynamic assignment}
\label{tb:route_result_dynamic}
\vspace{-0.3cm}
\begin{tabular}{|c|c|c|C{3.9cm}|c|c|c|c|}
\hline
Method &  $t^{ins}$(mins)    & $i$   & $path$   \\
\hline
\multirow{3}{*}{ \shortstack{Algorithms in \\ \cite{2019-UAV-route-optimize-fuel} }}  & \multirow{3}{*}{$40.5051$}    & \multirow{2}{*}{$1$} & B110$>$C214$>>$A106$>$B110     \\
\cline{4-4}
        				& &   & B110$>$E105$>$A411$>$B110 \\
\cline{3-4}
                         & & $2$ & A213$>$D101$>$A213          \\
\hline\hline
\multirow{2}{*}{ Algorithm \ref{ago:optimal-route} } & \multirow{2}{*}{$30.0741$}    & $1$ & B110$>$C214$>$A106$>$A411$>$B110     \\
\cline{3-4}
                         & & $2$ & A213$>$D101$>$E105$>$A213           \\
\hline
\end{tabular}
\vspace{-0.5cm}
\end{center}
\end{table}

\section{Conclusion}\label{sec:conclusion}

In this paper, we presented a framework of utilizing UAVs as the computing units and remote sensing units for offshore wind farms.
Then, we formulated two optimization problems to maximize the power generation of wind turbines and minimize the flight time for inspection.
To solve the formulated problems, we used wind forecasting as the input.
Conventional LSTM requires huge storage space and complex computation.
We utilized quantization and sparse matrix computation to address the issue of the limited resources on the UAVs.
With the forecasted result, we presented two algorithms to solve the optimization problems. 
We utilized real-world data to evaluate the proposed method.
With the proposed framework, wind turbines can reach close to the maximum power generation and reduce the inspection time by $25\%$.
%In the future work, we will explore to bring more function to UAVs such UAVs can provide more service in the wind farm.

\section{Acknowledgements}
This work was supported by Norwegian Research Council under Grants 275106 (LUCS project), 287412 (PACE project), and 267967 (SmartNEM project).

{\renewcommand{\baselinestretch}{1}
\begin{footnotesize}
\bibliographystyle{IEEEtran}
\bibliography{References_qRNN}
\end{footnotesize}}

\end{document}